\UseRawInputEncoding
\documentclass[aps,prb,superscriptaddress,longbibliography,showpacs,twocolumn]{revtex4-2} 
\usepackage{amsmath,amssymb}
\usepackage{graphicx}
\usepackage{xcolor}

\usepackage{bm}

\definecolor{LinkColor}{rgb}{0.256,0.439,0.588}
\usepackage{hyperref}
\hypersetup{
	pdfauthor={good guys},
	pdftitle={good title},
	colorlinks=true,
	citecolor=LinkColor,
	linkcolor=LinkColor,
	urlcolor=LinkColor
}

\newcommand{\beq} {\begin{equation}}
\newcommand{\eeq} {\end{equation}}
\newcommand{\bea} {\begin{eqnarray}}
\newcommand{\eea} {\end{eqnarray}}
\newcommand{\be} {\begin{equation}}
\newcommand{\ee} {\end{equation}}

\newcommand{\ket}[1]{\left\lvert#1\right\rangle}
\newcommand{\bra}[1]{\left\langle#1\right\rvert}

\begin{document}
\newcommand{\g}{g}
\title{Non-Hermitian Stark Many-Body Localization}

\author{Han-Ze Li}
\altaffiliation{The first two authors contributed equally.}
\affiliation{School of Physics and Optoelectronics, Xiangtan University, Xiangtan 411105, China}

\author{Xue-Jia Yu}
\altaffiliation{The first two authors contributed equally.}
\affiliation{Department of Physics, Fuzhou University, Fuzhou 350116, Fujian, China}

\author{Jian-Xin Zhong}
\email{jxzhong@xtu.edu.cn}
\affiliation{School of Physics and Optoelectronics, Xiangtan University, Xiangtan 411105, China}
\affiliation{Department of Physics, Shanghai University, Shanghai 200444, China}

\date{\today}

\begin{abstract}
Utilizing exact diagonalization (ED) techniques, we investigate a one-dimensional, non-reciprocal, interacting hard-core boson model under a Stark potential with tail curvature. By employing the non-zero imaginary eigenenergies ratio, half-chain entanglement entropy, and eigenstate instability, we numerically confirm that the critical points of spectral real-complex (RC) transition and many-body localization (MBL) phase transition are not identical, and an examination of the phase diagrams reveals that the spectral RC transition arises before the MBL phase transition, which suggests the existence of a novel non-MBL-driven spectral RC transition. These findings are quite unexpected, and they are entirely different from observations in disorder-driven interacting non-Hermitian systems. This work provides a useful reference for further research on phase transitions in disorder-free interacting non-Hermitian systems.
\end{abstract}
\maketitle

\section{INTRODUCTION}
\label{sec:introduction}
Non-Hermitian quantum systems have garnered a surge of research interest over the past two decades~\cite{1, 2, 3, 4, 5, 6, 7, 8, 9, 10, 11, 12, 13, 14, 15, 16, 17, 18, 19, 20, 21, 22, 23, 24, 25, 26, 27, 28, 29, 30, 31, 32}. This is due to their unique ability to host a range of novel quantum phase transitions (QPTs) without Hermitian counterparts, such as the spectral RC transitions, the topology phase transitions, and non-Hermitian skin effects, among others~\cite{PhysRevB.58.8384, RevModPhys.88.035002, PhysRevLett.121.136802, PhysRevB.106.144208, HATANO1998317, Weidemann2022, Wang2022,PhysRevLett.129.070401,Zhang2022, PhysRevB.97.121401, PhysRevB.104.195102, PhysRevB.99.201103}. The introduction of non-Hermiticity in quantum systems is typically achieved through the modulation of gain and loss in on-site energies or by manipulating non-reciprocal hopping. These two approaches exhibit distinct symmetries: the former maintains $\mathcal{PT}$-symmetry (referred to as NH-$\mathcal{PT}$ systems), while the latter aligns with time-reversal symmetry (NH-TRS systems). In NH-$\mathcal{PT}$ systems, Ref.~\cite{1} reports that where the on-site energies possess gain and loss, the eigenenergies are real if the system exhibits $\mathcal{PT}$-symmetry. Notably, the breaking of $\mathcal{PT}$-symmetry plays a crucial role in controlling the spectral RC transition observed in the eigenenergies. Conversely, NH-TRS systems exhibit another spectral RC transition induced by MBL. This was initially addressed in Ref.~\cite{PhysRevLett.77.570, PhysRevLett.123.090603}. Further investigations by Ref.~\cite{PhysRevB.102.064206} unveiled a comparable MBL-driven spectral RC transition in one-dimensional interacting NH-TRS systems subjected to a quasi-periodic potential. Remarkably, the critical points associated with the spectral RC transition and the MBL phase transition coincide in the thermodynamic limit~\cite{113} for both random and quasi-periodic potentials~\cite{PhysRevLett.123.090603, PhysRevB.102.064206}. Despite the fact that disorder-induced systems, encompassing both random and quasi-periodic scenarios, do not belong to the same universality class from the perspective of the renormalization group (RG), disorder emerges as the overarching factor inducing MBL in both cases.

However, it is important to note that disorder is not the sole mechanism leading to MBL (Anderson Localization~\cite{PhysRev.109.1492}, AL) in many-body (single-particle) systems. 
In the context of single-particle scenarios, the application of a gradient external electric field can give rise to states reminiscent of AL, exhibiting an exponential decay of localization in a system initially in an extended state. This phenomenon is known as Wannier-Stark localization~\cite{PhysRevB.36.7353}, with the applied external electric field referred to as the Stark potential. Similarly, in the realm of many-body systems, the behavior of Stark many-body localization (SMBL)~\cite{PhysRevLett.122.040606}, akin to MBL, has been observed in both the static and dynamic responses of systems subjected to a Stark potential. Notably, compared to disorder-induced systems, these disorder-free systems exhibit cleaner and simpler characteristics, as evidenced by experimental observations~\cite{PhysRevLett.127.240502, Morong2021} and numerical simulations~\cite{doi:10.1073/pnas.1819316116, PhysRevB.104.205122}. Consequently, they provide a fresh platform for the exploration of MBL and offer promising prospects for a range of applications. Recent experimental realizations in ion traps~\cite{Morong2021} and superconducting circuits~\cite{PhysRevLett.127.240502} have further demonstrated the feasibility of studying SMBL, underscoring the potential and exciting avenues for future exploration in this field.

Naturally, several significant questions arise: Can MBL phase transitions and MBL-driven spectral RC transitions occur in disorder-free non-Hermitian many-body systems?  Given the notable advantages offered by SMBL systems, including their suitability for experimental observations and numerical simulations, as well as their capacity to host novel QPTs, addressing this question is of significant interest and not to be underestimated.

To address the aforementioned question, in this work, we investigate the spectral RC transition and MBL phase transition in a one-dimensional interacting NH-TRS hard-core bosonic chain model subjected to a Stark potential with a tail curvature. Through the large-scale ED simulations, we observe the coexistence of the MBL and spectral RC transition in the phase diagram. However, we suggest that the critical points associated with the spectral RC transition and MBL phase transition might be distinct in the thermodynamic limit. Notably, our analysis reveals the presence of a novel spectral RC transition that is not triggered from MBL. These findings present a striking departure from those observed in disorder NH-TRS many-body systems.

The rest of the paper is organized as follows: Section \ref{sec:model} contains an overview of the interacting NH-TRS model under Stark potential and the numerical methods employed. Section \ref{sec:results} presents the numerical results, identifies the physical quantities necessary for detecting spectral RC transition and MBL phase transition, and provides the phase diagram of NH-SMBL. The resolution is described in Section \ref{sec:con}. Additional data supporting our numerical calculations can be found in the Appendixes.

\section{MODEL AND METHOD}
\label{sec:model}
We consider a one-dimensional, NH-TRS hard-core boson chain with a Stark potential. This model consists of $L$ sites and is represented as follows, under periodic boundary conditions (PBC):
\begin{align}
    \hat{H}=\sum_{j=1}^{L}\left[-t(e^{-g}\hat{b}^\dagger_{j+1}\hat{b}_j+e^{g}\hat{b}^\dagger_j \hat{b}_{j+1})+U\hat{n}_{j}\hat{n}_{j+1}+\Delta_j \hat{n}_j \right ].\label{eq1}
\end{align}
\noindent In the given context, $L$ denotes the length of the lattice. The terms $t_L \equiv te^{g}$ and $t_R \equiv te^{-g}$ represent the non-reciprocal hopping strengths towards the left and right respectively, where $g$ is the strength of non-reciprocal hopping. The parameter $U$ characterizes the strength of the nearest-neighbor interaction. The term $\Delta_j$ embodies the Stark potential with tail curvature, which is given as follows:
\begin{align}
    \Delta_j \equiv -\gamma j + \alpha (j/L)^2.
\end{align}
\noindent Here, $\gamma$ symbolizes the strength of the Stark potential and $\alpha$ signifies the curvature of its tail. The operator $\hat{b}_j$ and $\hat{b}_j^\dagger$ denote the annihilation and creation operations of hard-core boson at the site $j$, respectively. They conform to the commutation relation $[\hat{b}_k,\hat{b}_l^\dagger]=\delta_{kl}$. The number operator for particles is denoted as $\hat{n}_j\equiv \hat{b}_j^\dagger \hat{b}_{j}$, signifying the count of particles at site $j$.

In this model, we highlight several key features: (a) In the case where $U=0$ and $g=0$, the model reverts to a Hermitian, single-particle scenario. Under these conditions, a Stark potential can induce what is known as Wannier-Stark localization. (b) The non-Hermitian setting of our model is pertinent to the scenario of continuously monitored quantum many-body systems. Our focus lies on single quantum trajectories without quantum jumps, that is, the post-selection of pure states as the outcome of the measurement. This approach provides a contrast to the Gorini-Kossakowski-Sudarshan-Lindblad equation methods for open system dynamics \cite{PhysRevLett.116.160401, PhysRevX.7.011034}, which yield average results. (c) The tail curvature parameter $\alpha$ confers stability to the MBL properties of systems under Stark potentials. For more details, please refer to Appendix ~\ref{sec:s3}.

In this paper, we use the ED method, with the aid of the QuSpin package \cite{10.21468/SciPostPhys.7.2.020}, to numerically derive the solutions pertaining to Eq. (\ref{eq1}). The parameters of the model are chosen as follows: $t=1.0$, $U=1.0$, $\alpha=0.5$, and $g=0.1$. A fixed particle-number subspace with $M = L/2$ particles is considered, which corresponds to a half-filled system. We assert that the spectral RC transitions in Eq. (\ref{eq1}) are robust against changes in the boundary conditions. We have placed the numerical results and discussions for open boundary conditions (OBCs) in Appendix ~\ref{sec:S2}.

\section{NUMERICAL RESULTS}
\label{sec:results}
\subsection{Spectral RC transitions}

Eigenenergies with a nonzero imaginary part ratio serve as a robust probe for detecting spectral RC transitions throughout the entire energy spectrum \cite{PhysRevLett.123.090603}. It is defined across the whole spectrum as
\begin{align}
    f _ {\rm I m } = D _ {\rm I m } / D.\label{eq6}
\end{align}
Here, $D_{\rm Im}$ represents the number of eigenenergies with nonzero imaginary components. To remove potential inaccuracies arising from numerical techniques, we define eigenenergies $E_{\rm \alpha}$ to have non-zero imaginary parts when ${\rm Im}(E_{\alpha}) \gg C$ ($C=10^{-13}$). Simultaneously, $D$ denotes the total number of eigenenergies. If all eigenenergies are purely real, it corresponds to $f_{\rm Im} = 0$, while in the extreme case where all eigenenergies are complex, this occurs at $f_{\rm Im} = 1$. It's important to note that the critical point and critical exponent in this case differ substantially from those of disorder-driven systems, suggesting that they are not part of the same universal class of criticality.

\begin{figure}[htbp]
    \centering
    \includegraphics[width=0.48\textwidth]{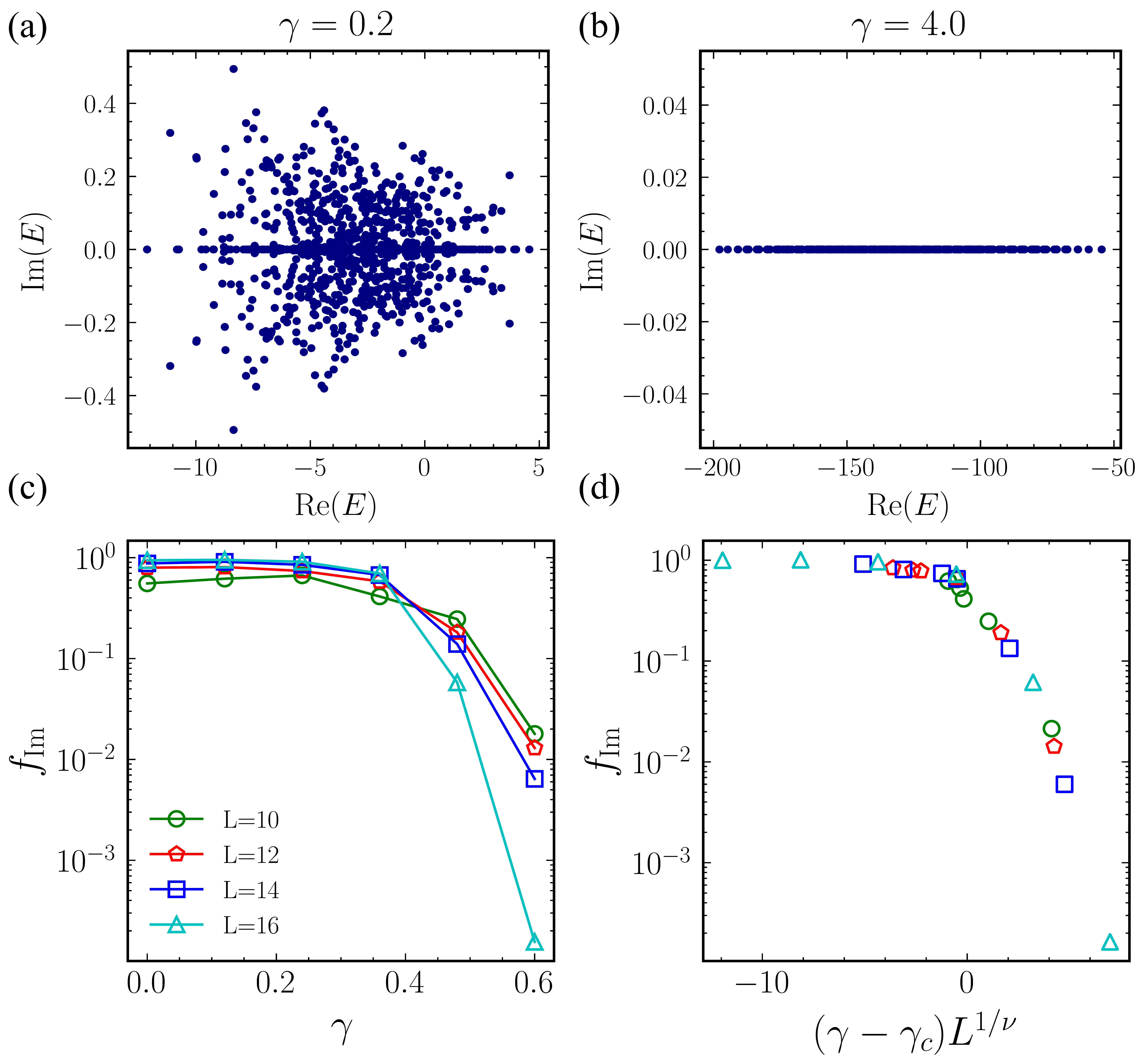}
    \caption{(Color online) Spectral RC transition. The eigenenergies of Eq.(\ref{eq1}) for (a) $\gamma=0.2$ and (b) $\gamma=4.0$ are shown. The parameters are $L=12$ (lattice size), $g=0.1$ (non-Hermitian strength), and $U=1.0$ (interaction strength). (c) Presents results from Eq.(\ref{eq6}) for lattice sizes $L=10, 12, 14, 16$. (d) Data points are fitted using the scale function $(\gamma-\gamma^{f}_c)L^{1/\nu}$, yielding a critical point at $\gamma^{f}_c \approx 0.42\pm0.15$ and a critical exponent $\nu \approx 0.78\pm0.1$.}
    \label{F1}
\end{figure}

The spectral RC transition of eigenenergies at size $L=12$ under $\gamma=0.2$ and $\gamma=4.0$ is depicted in FIG. \ref{F1}(a) and FIG. \ref{F1}(b) respectively. Owing to the TRS, all eigenenergies with imaginary parts are symmetrically distributed along the real axis. Notably, when $\gamma=0.2$, there are a greater number of eigenenergies with non-zero imaginary parts. Conversely, in a deeper MBL region, specifically when $\gamma=4.0$, almost all eigenvalues fall on the real axis. FIG. \ref{F1}(c) illustrates a critical point, $\gamma^{f}_c \approx 0.42\pm0.15$, beyond which the value of $f_{\rm Im}$ decreases as $\gamma$ increases.  We confirm that the critical scaling collapse as a function of $(\gamma-\gamma^{f}_c)L^{1/\nu}$, as utilized in FIG. \ref{F1}(d), reveals a critical exponent of $\nu \approx 0.78\pm0.10$.

\subsection{MBL phase transitions}

Level-spacing statistics provide an effective method to probe the energy spectra of quantum systems, revealing characteristics of the Hamiltonian such as integrable-chaotic spectra, QPTs, and symmetry-breaking phenomena. Nevertheless, due to differences in Hermitian properties, the level-spacing statistical analysis applied in Hermitian systems cannot be directly utilized for non-Hermitian systems \cite{PhysRevX.12.021040, PhysRevX.12.011006, Haake2010}. As a result of changes in matrix symmetries, the 10-fold symmetry classification in Hermitian systems expands to a 38-fold classification in the non-Hermitian realm \cite{PhysRevX.9.041015, PhysRevB.55.1142, PhysRevB.99.235112}. Given that eigenvalues in non-Hermitian systems are points distributed across a two-dimensional (2D) complex plane, the unfolding nearest-neighbor level-spacing $s_\alpha$ serves as the statistical data.

First, we stipulate that the minimum distance in the complex plane for the eigenvalue $E_m$ is denoted as
\begin{align}
    s_{1,m}\equiv \min_{l}|E_m -E_l|.
\end{align}
Then, we define a local mean density of eigenvalues, denoted as $\bar{\rho}$, which is 
\begin{align}
    \bar{\rho}=n/(\pi s^2_{n,m}),
\end{align}
where $n$ is a larger than unitary (approximately $30$), and $s_{n,m}$ represents the $n$th-nearest-neighbor distance from $E_{\alpha}$. Finally, we obtain the unfolding normalized nearest-neighbor level spacing, 
\begin{align}
    s_{\alpha}=s_{1,m}\sqrt{\bar{\rho}}.
\end{align}
We will use it for the calculation of level spacing statistics in the following.

Contrasting with standard level statistics methods, in the realm of complex eigenvalue space, concerns over the influence of local eigenvalue density on level-spacing become obsolete \cite{Haake2010, PhysRevLett.123.090603,PhysRevResearch.2.023286}. Now, we can do the statistics of unfolding normalized nearest-neighbour spacing using $s_\alpha$.

In the delocalization phase, the non-Hermitian probability distribution $p(s)$ follows the Ginibre distribution $P^{\rm C}_{\rm Gin}(s)=cp(cs)$. This distribution characterizes an ensemble of non-Hermitian Gaussian random matrices \cite{Haake2010}. The specific form of this distribution is given by:
\begin{align}
    p(s)= \lim _{N \rightarrow \infty}\left[ \prod _{n=1}^{N-1}e_{n}(s^{2})e^{-s^{2}}\sum _{n=1}^{N-1}\frac{2s^{2n+1}}{n!e_{n}(s^{2})}\right]
\end{align}
where,
\begin{align}
    e_{n}(x)= \sum _{m=0}^{n}\frac{x^{m}}{m!}
\end{align}
and
\begin{align}
    c= \int _{0}^{\infty}dssp(s)=1.1429\cdots.
\end{align}
For the MBL phase, with the eigenenergies being localized on the real axis during this stage, the level-spacing statistics follow a Poisson distribution \cite{Haake2010}, denoted as 
\begin{align}
    P_{\rm Po}^{\rm R}(s)=e^{-s}.
\end{align}

\begin{figure*}[htbp]
    \centering
    \includegraphics[width=1\textwidth]{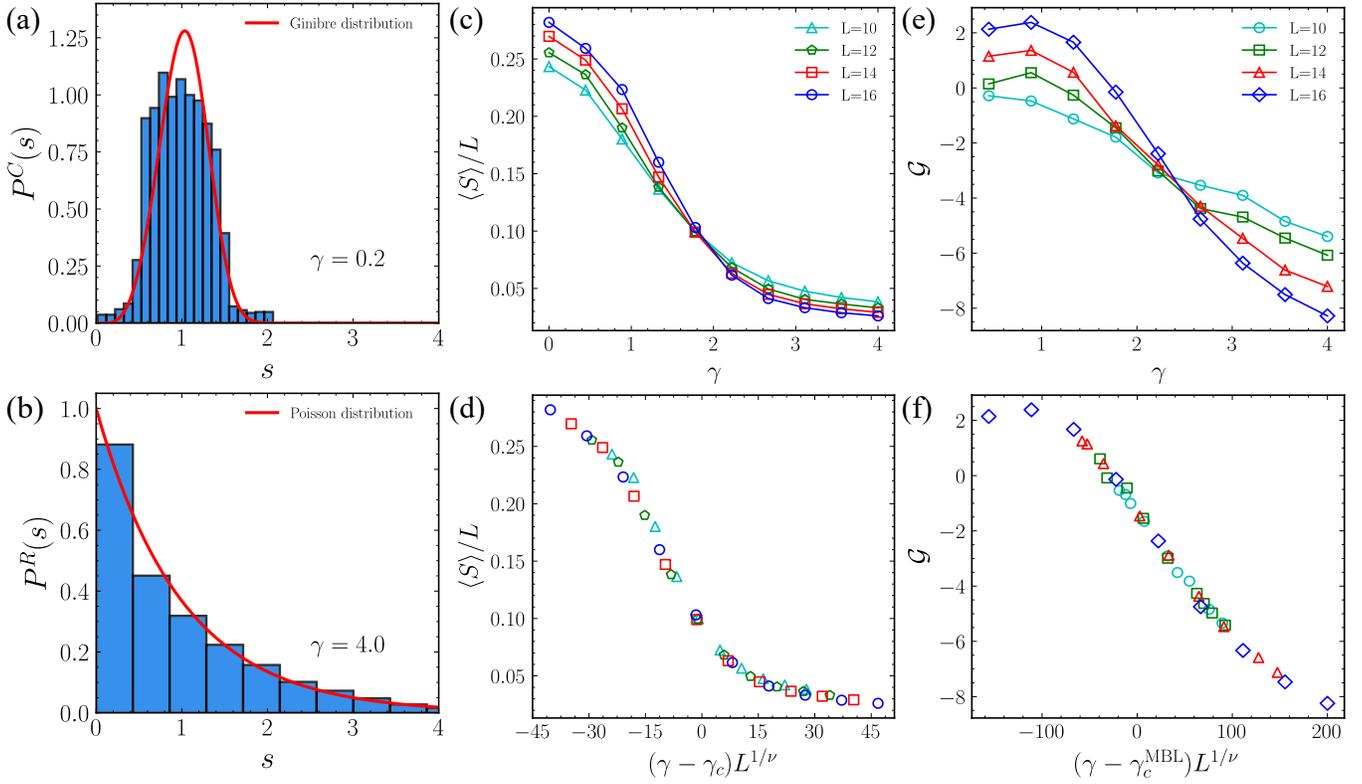}
    \caption{(Color online) The MBL phase transition. (a) Depicts the level-spacing distribution for $\gamma=0.2$, with the red solid line indicating the Ginibre distribution. (b) Displays the level-spacing distribution for $\gamma=4.0$, with the red solid line corresponding to the Poisson distribution. (c) Illustrates the relationship between the average entanglement entropy per system size $\langle S \rangle/L$ of selected right eigenstates within a central range of the real part of the spectrum (specifically those within $\pm 4 \%$ of the middle), and the Stark potential strength $\gamma$, across different system sizes $L$. (d) Offers a data collapse fit to the data from (c) using the scaling function $(\gamma-\gamma_c)L^{1/\nu}$, yielding a critical point $\gamma_c \approx 1.92\pm0.24$ and critical exponent $\nu \approx 0.90\pm 0.10$. (e) Portrays the dependence of eigenstate instability on the Stark gradient potential $\gamma$ for different sizes $L$. (f) Presents a fit of the data in (e), with the approximated critical point being $\gamma^{\rm MBL}_c \approx 2.17\pm0.10$ and the approximated critical exponent being $\nu \approx 0.63 \pm 0.11$.}
    \label{f1}
\end{figure*}

In analyzing the level-spacing statistics of non-Hermitian systems, we concentrate on eigenenergies situated at the center of the spectrum. We include both real and imaginary parts within a $\pm 10 \%$ range, as specified in Ref.~\cite{PhysRevLett.123.090603}. As depicted in FIG. \ref{f1}(a), the distribution conforms to the Ginibre distribution when the Stark potential strength is set to $\gamma=0.2$. Yet, when the Stark potential strength is increased to $\gamma=4.0$, the distribution transitions to follow the Poisson distribution, as shown in FIG. \ref{f1}(b). This finding suggests that the system undergoes a MBL phase transition in response to variations in the Stark potential strength $\gamma$.

Upon confirming the existence of an MBL phase transition in the system, we turn to get the critical information. Firstly, we consider the static half-chain entanglement entropy, restricting our calculations to normalized right eigenstates $\ket{\varepsilon^r_n}$ $i. e.$, $\langle\varepsilon^r_n|\varepsilon^r_n\rangle=1$.
The specific form is as follows: 
\begin{align}
    S_n=-{\rm Tr}[\rho^n_{L/2}\ln{\rho^n_{L/2}}],
\end{align}
Here, the reduced density matrix for the half chain, obtained by performing a trace over half of the system,
\begin{align}
    \rho^n_{L/2} = {\rm Tr}_{L/2}[\ket{\varepsilon_n^r}\bra{\varepsilon_n^r}].
\end{align}

As demonstrated in FIG.~\ref{f1}(c) and (d), these figures depict the relationship between the average entanglement entropy per system size $\langle S \rangle/L$ for selected right eigenstates from the middle of the real part spectrum (specifically within a $\pm 4 \%$ range) and the Stark potential strength $\gamma$, as a function of system size $L$. In FIG. \ref{f1}(c), it is clearly visible that there is a transition from volume law to area law for the entanglement entropy around the critical point. We have set the function form for the critical scaling collapse as $(\gamma-\gamma_c)L^{1/\nu}$. FIG. \ref{f1}(d) presents the rescaled curve, from which we have identified the range of the critical point as $\gamma_c \approx 1.92\pm0.24$, and the critical exponent as $\nu \approx 0.90\pm 0.10$. 

However, for interacting NH-TRS Hamiltonians, the choice of right eigenstates can significantly impact the quantitative results of critical values \cite{110}. A more effective method to understand MBL phase transitions is through the examination of eigenstate instability. A defining feature of localized eigenstates is their robustness against local disturbances \cite{PhysRevX.5.041047, PhysRevLett.118.266601}. In the realm of non-Hermitian quantum many-body systems that uphold TRS, localization prompts the clustering of imaginary eigenenergies onto the real axis. Therefore, introducing localized perturbations to examine the stability of eigenstates against such perturbations becomes critical. This approach serves as a way to identify MBL phase transitions in Hermitian systems. Within the non-Hermitian framework, we can define a stability index for the eigenstates as follows:
\begin{align}
    \mathcal{G}= \ln \frac{|\bra{\varepsilon^l_{a+1}}\hat{V}_{NH}\ket{\varepsilon^r_{a}}|}{|\varepsilon'_{a+1}-\varepsilon'_{a}|}
\end{align}
Here, $\ket{\varepsilon_{a}^l}$ and $\ket{\varepsilon_{a}^r}$ represent the left and right eigenstates of the non-Hermitian Hamiltonian $\hat{H}$, respectively, which satisfy $\langle\varepsilon_{a}^l|\varepsilon_{b}^r\rangle=\delta_{ab}$. The perturbation term is given by $\hat{V}_{\rm NH}=\hat{b}^{\dagger}_{j}\hat{b}_{j+1}$. We obtain $\varepsilon'_{a}=\varepsilon_{a}+\bra{\varepsilon_{a}^l}\hat{V}_{\rm NH}\ket{\varepsilon_{a}^r}$. And the set of $\{\varepsilon'_a\}$ is also positive and sorted in ascending order. For the ergodic phase, $\partial G / \partial L >0$ is valid \cite{PhysRevE.91.012144, PhysRevLett.111.050403}, whereas for the localization phase, $\partial G / \partial L <0$ is observed \cite{PhysRevLett.117.027201}.

For the choice of the local operator $\hat{V}_{\rm NH}$, we follow the same form as described in Ref.~\cite{PhysRevLett.123.090603}. As depicted in FIG. \ref{f1}(e), the results showcase the dependence of the eigenstate stability, $\mathcal{G}$, on $\gamma$. Before the critical point $\gamma^{\rm MBL}_c$, $\mathcal{G}$ increases with the increase in size $L$, indicating an ergodic phase represented by $\mathcal{G} \sim \zeta L$. However, after the critical point $\gamma^{\rm MBL}_c$, $\mathcal{G}$ decreases as the size $L$ grows, signifying a MBL phase represented by $\mathcal{G} \sim -\eta L$. Here, $\eta$ and $\zeta$ represent coefficients that are greater than zero. As illustrated in FIG. \ref{f1}(f), we have identified the critical value as $\gamma^{\rm MBL}_c = 2.17\pm0.10$ and the critical exponent is $\nu \approx 0.63 \pm 0.11$.

The observed numerical behavior contrasts significantly with what is typically seen in disorder-induced interacting NH-TRS systems, where the spectral transition and MBL transition usually occur concurrently. However, under a Stark potential with tail curvature, the interacting NH-TRS system initially undergoes a spectral RC transition, followed by an MBL phase transition. This sequence does not contradict the concept that MBL can induce coalescence of imaginary energies. It is distinctly different from the behavior observed in disorder-induced interacting NH-TRS systems. However, the fact that MBL occurs after the spectral RC transition suggests that the Stark potential begins to influence the spectral RC transition even before the onset of the MBL phase transition. This directs the system through a previously unexplored intermediate phase, where the spectrum is real and ergodic, and the spectral RC transition is not principally driven by the MBL. 

We choose not to delve into an extensive discussion about this observation in the current work, but we anticipate a systematic investigation of this intriguing issue in future studies. Moreover, although the topological phase transition is not closely related to the primary focus of our study, we do observe its presence in the interacting Stark system with non-reciprocal hopping. However, the behavior of topological phase transitions in non-Hermitian interacting systems is complex. Therefore, for the benefit of the readers, we have included the numerically observed results obtained using ED in Appendix ~\ref{sec:S1}.

\subsection{Phase diagram}

\begin{figure}[htbp]
    \centering
    \includegraphics[width=0.48\textwidth]{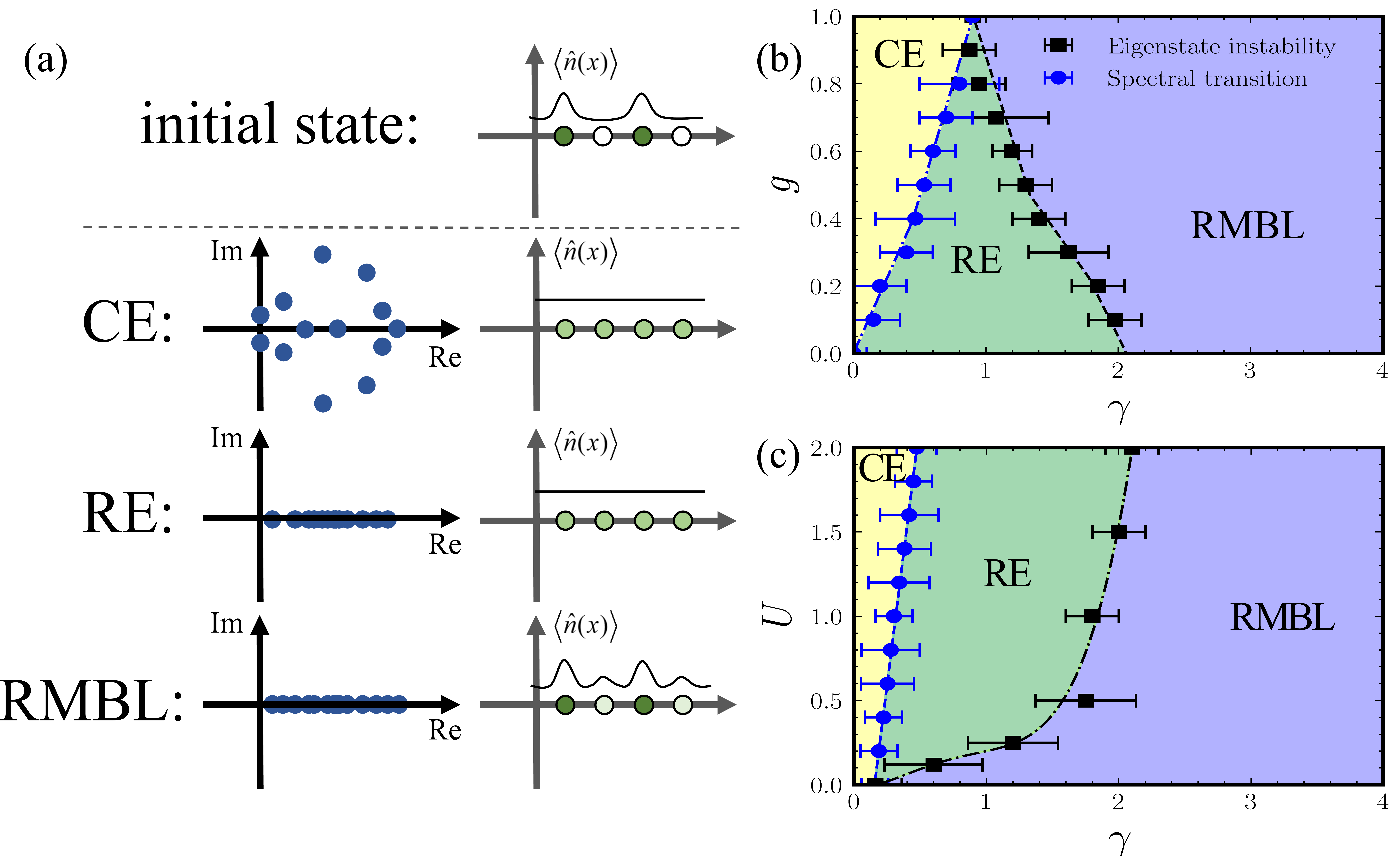}
    \caption{(Color online) Schematic of the three phases under Stark potential (a), phase diagrams of non-reciprocal hopping strength $g$ and Stark potential strength $\gamma$ (b), and interaction strength $U$ and Stark potential $\gamma$ (c). In (a), we denote by CE phase in which the spectrum is complex and occupies an ergodic phase. The term RE is used for a phase featuring a real spectrum also in an ergodic phase. Meanwhile, the acronym RMBL characterizes a phase with a real spectrum, but in a MBL phase. In (b) and (c), the blue markers stand for the numerical outcomes procured through ED, with the associated error bars calculated based on each respective data point. The blue dotted line, a result of fitting efforts, signifies the spectral transition boundaries. The black markers, on the other hand, are indicative of eigenstate instability outcomes. The black dashed line, another fitting result, serves to demarcate the border between MBL and ergodic phase.}
    \label{f0}
\end{figure}

The phase diagrams and transitions \cite{112} for Stark potential strength $\gamma$ in Eq. [\ref{eq1}] are obtained by performing ED simulations with sizes $L=10, 12, 14, 16$, as shown in FIG. \ref{f0}. FIG. \ref{f0}(a) provides a schematic of the three phases under the Stark potential: complex spectrum and ergodic (CE), real spectrum and ergodic (RE), and real spectrum with many-body localization (RMBL). Here, $\langle \hat{n}(x) \rangle$ represents the average number of particles at different positions. In FIG. \ref{f0}(b), we fix the interaction strength at $U=1.0$. The blue dashed line on the left represents the increase of eigenenergies with a nonzero imaginary part ratio of eigenvalues $f_{\rm Im}$ as the size $L$ increases. Conversely, on the right, it corresponds to the decrease of $f_{\rm Im}$ with increasing $L$. The black dotted line on the left signifies the growth of the eigenstate stability $\mathcal{G}$ with increasing $L$, whereas, on the right, it signifies the decrease of $\mathcal{G}$ as $L$ expands. The CE phase (the yellow region), characterized by a complex spectrum, depicts an area where both $\mathcal{G}$ and $f_{\rm Im}$ increase with the enlargement of dimension $L$, signifying that the system resides in an ergodic phase. The RE phase (the green region), marked by a real spectrum, represents an area where $\mathcal{G}$ decreases and $f_{\rm Im}$ increases with the expansion of dimension $L$, indicating that the system inhabits an ergodic regime. The RMBL phase (the purple region), distinguished by a real spectrum, corresponds to an area where both $\mathcal{G}$ and $f_{\rm Im}$ decrease as dimension $L$ expands, denoting that the system is in MBL regime.

\begin{figure*}[htbp]
    \centering
    \includegraphics[width=1\textwidth]{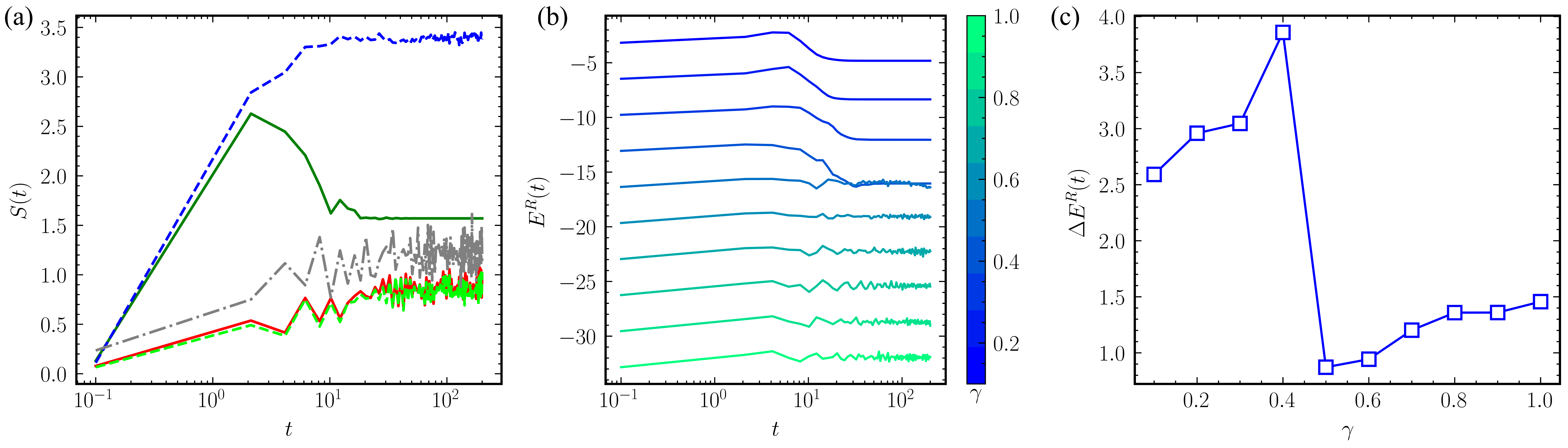}
    \caption{(Color online) Dynamics of the transitions with the initial state $\ket{\psi_0}=\ket{10101\cdots}$. (a) The dynamics of the half-chain entanglement entropy $S(t)$ for $U=1.0, g=0.1, \gamma=0.2$ (green solid line), $U=1.0, g=0, \gamma=0.2$ (blue dashed line), $U=1.0, g=0.1, \gamma=4.0$ (red solid line), $U=1.0, g=0, \gamma=4.0$ (lime dashed line), and $U=1.0, g=0.8, \gamma=0.2$ (blue dashed line). (b) The evolution of $E^{R}(t)$ for $\gamma$ ranging from $0.1$ to $1.0$. The color bar shows the values of different $\gamma$. (c) $\Delta E^{R}(t)$ as a function of $\gamma$.}
    \label{f4}
\end{figure*}

In FIG.~\ref{f0}(c), with non-reciprocal strength $g$ set at $g=0.1$, we identify three regions. CE is characterized by an increase in both $\mathcal{G}$ and $f_{\rm Im}$ as the size $L$ expands. Conversely, RE depicts a scenario where $\mathcal{G}$ diminishes with an enlarging $L$, while $f_{\rm Im}$ continues to rise. Lastly, RMBL signifies an area where both $\mathcal{G}$ and $f_{\rm Im}$ decrease in response to the growth of $L$. The error bars are deduced from the shifts at the transition points across various system sizes. In the non-interacting limit where $U=0$, the transitions coincide, aligning with the conclusions of Ref.~\cite{PhysRevB.56.R4333}. This alignment implies that the spectral transition behavior demonstrated by the MBL phase transition is consistent with RC transition. Nevertheless, in the presence of interactions, these two transitions diverge as $U$ increases. This divergence suggests that an intermediate phase arises as $U$ intensifies, a phase solely observed in non-Hermitian SMBL systems.

\subsection{Dynamics of transitions}

We now focus to the dynamical behavior of the system in response to phase transitions. The dynamics of entanglement entropy $S(t)$.
In a non-Hermitian quantum system with no quantum jump \cite{PhysRevX.7.021013}, we can evolve a given initial state $\ket{\psi_0}=\ket{10101\cdots}$ as 
\begin{align}
    \ket{\psi(t)}=\frac{e^{-i\hat{H}t}\ket{\psi_0}}{||e^{-i\hat{H}t}\ket{\psi_0}||}.
\end{align}
Here, $||e^{-i\hat{H}t}\ket{\psi_0}||=\sqrt{\bra{\psi_0}e^{i\hat{H}^{\dagger}t}e^{-i\hat{H}t}\ket{\psi_0}}$. The time-dependent reduced density matrix of half-chain is 
\begin{align}
    \rho_{L/2}(t)={\rm Tr}_{L/2}[\ket{\psi(t)}\bra{\psi(t)}].
\end{align}
The dynamics of entanglement entropy $S(t)$ is given by
\begin{align}
    S(t)=-{\rm Tr}[\rho_{L/2}(t)\ln\rho_{L/2}(t)].
\end{align}

 As illustrated in FIG. \ref{f4}(a), we examine the dynamic trajectory of the half-chain entanglement entropy in both the Hermitian boundary case ($g=0$) and the non-Hermitian scenario ($g=0.1$). We do this for two distinct phases: an ergodic phase with $\gamma=0.2<\gamma^{\rm MBL}_c$ and a localized phase with $\gamma=4.0>\gamma^{\rm MBL}_c$. In FIG. \ref{f4}(a), during the ergodic phase, a notable difference emerges between the Hermitian case $g=0$ (represented by the blue dashed line) and the non-Hermitian case $g=0.1$ (represented by the green solid line). For $g=0$, the entanglement entropy $S(t)$ initially exhibits linear growth, then stabilizes around $S(t)\approx 3.4$. In contrast, for $g=0.1$, $S(t)$ initially grows linearly until $t\approx 20$, then decreases and eventually stabilizes at $S(t)\approx 1.5$. Conversely, in the MBL phase ($\gamma=4.0$), both $g=0$ and $g=0.1$ exhibit slow logarithmic growth of $S(t)$. This observation is consistent with the dynamics of entanglement entropy in disorder-driven MBL systems \cite{PhysRevLett.123.090603,PhysRevB.102.064206}. In the MBL phase, the influence of $g$ merely results in an overall upward shift of $S(t)$, as indicated by the gray dotted line. The distinctive behavior between the Hermitian and non-Hermitian cases in the ergodic phase can be attributed to the eigenstates of complex eigenvalues. For non-Hermitian systems in the ergodic phase, we observe an unusual decrease following a certain growth in the half-chain entanglement entropy. As the system enters the ergodic phase, the number of complex eigenenergies increases, leading to a gradual reduction in entanglement over time. Ultimately, the entanglement entropy stabilizes at a specific value, denoting the influence of non-Hermiticity on entanglement dynamics in the ergodic phase. However, in the MBL phase, in the absence of eigenstates with non-zero imaginary parts of their eigenenergies, both the Hermitian and non-Hermitian cases exhibit similar behavior.
 
When exposed to extreme disorder or intense Stark potentials, highly excited eigenenergies transition to purely real values. This transition induces a significant shift in the dynamic properties of the system. A primary measure of this transformative dynamic behavior is reflected in the evolution of the real part of the energy within the system \cite{PhysRevLett.123.090603},
\begin{align}
    E^{R}(t)={\rm Re} [\bra{\psi(t)}\hat{H}\ket{\psi(t)}].
\end{align}

FIG. \ref{f4}(b) depicts the response of the real part of the eigenenergy, denoted as $E^R(t)$, to the phase transition. For $\gamma \leq \gamma_c^{f}$, a decrease in $E^R(t)$ around $t=10$ followed by stabilization is observable. However, this behavior is absent when $\gamma>\gamma_c^{f}$. Additionally, we examined the fluctuation of $E^{R}(t)$, as presented in FIG. \ref{f4}(c), defined as $\Delta E^{R}(t)=|\max{E^{R}(t)}-\min{E^{R}(t)}|$, across different $\gamma$ parameters. Notably, a sharp drop in $\Delta E^{R}(t)$ starting around $\gamma=0.4$ was observed. This feature echoes the vicinity of $\gamma_c^{f}$ in static situations, signaling the onset of the spectral RC transition.

\section{CONCLUSION AND OUTLOOK}
\label{sec:con}
In conclusion, through ED simulations, we have examined the critical behavior of spectral RC transition and MBL phase transition in a one-dimensional interacting NH-TRS hard-core boson chain subjected to Stark potential. By employing the ratio of non-zero imaginary eigenenergies and eigenstate instability as indicators, we've constructed a ground state phase diagram spanning CE, RE, and RMBL phases. As the Stark potential intensifies, we identified the critical point for the MBL phase transition at $\gamma^{\rm MBL}_{c}\approx 1.92\pm0.24$, and for the spectral RC transition at $\gamma^f_{c}\approx 0.42\pm0.15$. Moreover, we've analyzed the dynamical behavior of both the real part of the eigenenergy and the entanglement entropy. Unexpectedly, our study revealed that in an interacting NH-TRS system with Stark potential, the critical points for the spectral RC transition and MBL phase transition suggest the possibility of not being identical at the thermodynamic limit. Most notably, Upon evaluating the phase diagram, we find that the spectral RC transition occurs before the MBL phase transition. This suggests the existence of a new non-MBL-driven mechanism for spectral RC transitions in interacting NH-TRS systems with Stark potential. Such a discovery sharply contrasts with disordered interacting NH-TRS systems, where the spectral RC transition and MBL phase transition suggest a potential simultaneous occurrence in the thermodynamic limit. (The rigorous stability of the MBL phase in the thermodynamic limit is a subject of ongoing debate~\cite{PhysRevE.102.062144, PhysRevB.102.064207}. Therefore, based on our numerical results from finite-size systems, we can only suggest the potential existence of the MBL phase in the thermodynamic limit based on observed trends.) The theoretical foundation of this finding could be further elucidated in future work. Our discoveries open up a new avenue for the exploration of a novel non-equilibrium QPT in disorder-free interacting non-Hermitian quantum systems.

\emph{Note added:} A follow-up study titled "From Ergodicity to Many-Body Localization in a One-Dimensional Interacting Non-Hermitian Stark System" [\emph{arXiv:2305.13636}] \cite{513} also finds alignment with our results in a similar model utilizing PBCs. This serves to further validate and underline the robustness of our findings within non-Hermitian many-body systems under the influence of a Stark potential.

\begin{acknowledgments}
 We thank Professor Ching Hua Lee from the National University of Singapore (NUS) for the fruitful discussions. We also appreciate the discussions and support from Professor Lijun Meng at Xiangtan University. Additionally, we extend our gratitude to Jie Liu, Chao Wang, Dan Mu, Quan Gao, and Jiamei Han for their support to this study. We gratefully acknowledge the National Natural Science Foundation of China (Grant No. 11874316), the National Basic Research Program of China (Grant No. 2015CB921103) and the Program for Changjiang Scholars and Innovative Research Team in University (Grant No. IRT13093).
\end{acknowledgments}

\bibliography{apssamp}

\begin{thebibliography}{79}
\bibitem [{113()}]{113}%
  \BibitemOpen
  \href@noop {} {}\bibinfo {note} {Note that the rigorous stability of MBL in
  the thermodynamic limit is a topic of ongoing debate.}\BibitemShut {Stop}%
\bibitem [{110()}]{110}%
  \BibitemOpen
  \href@noop {} {}\bibinfo {note} {If the spectral sampling range ($\pm 4\%$)
  is expanded, a slight forward shift of the critical point would be observed.
  This is predominantly because increasing the spectral sampling range entails
  the inclusion of more eigenenergy levels with nonzero imaginary parts. These
  levels causing a drop in entanglement entropy at smaller $\gamma$ values and
  resulting in a forward shift of the critical point.}\BibitemShut {Stop}%
\end{thebibliography}

\newpage
\onecolumngrid

\appendix

\section{TOPOLOGICAL PHASE TRANSITIONS}
\label{sec:S1}

In one-dimensional interacting non-Hermitian systems, the winding number, defined on the complex plane, functions as the effective topological invariant. Gauge transformations $\hat{b}_{j}\rightarrow e^{i \frac{\Phi}{L}j}\hat{b}_{j}$ and $\hat{b}^{\dagger}_{j}\rightarrow e^{-i \frac{\Phi}{L}j}\hat{b}_{j}$ are introduced alongside parameter $\Phi$ (commonly interpreted as magnetic flux) to transform the model Hamiltonian into an effective form, 

\begin{figure*}[htbp]
    \centering
    \includegraphics[width=1\textwidth]{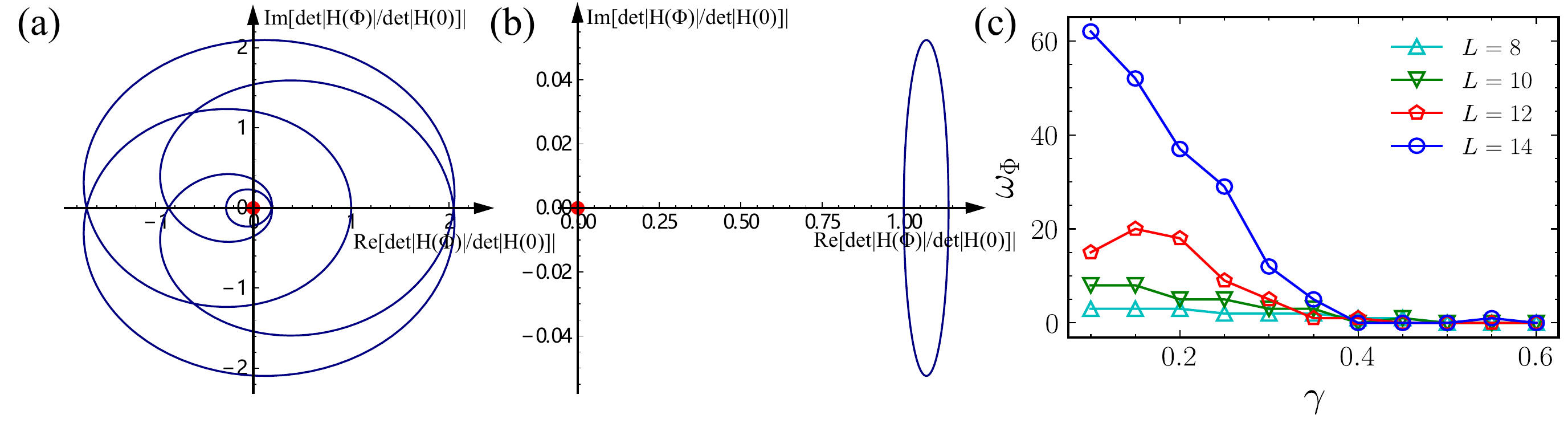}
    \caption{(Color online) The dependence of ${\rm det}H(\Phi)/|{\rm det}H(0)|$ on $\omega_{\Phi}$, with $\omega_{\Phi}$ varying from $0$ to $2\pi$, is shown for (a) $\gamma=0.3$ and $\gamma=0.45$. The red points represent the origin point $E_B=0$. (c) The function of the winding number $\omega_{\Phi}$ and $\gamma$ is shown for $L=10, 12, 14, 16$. As $\gamma$ increases, $\omega_{\Phi}$ gradually decreases, eventually dropping to $0$. The topological phase transition critical point is at $\gamma_c^{T}\approx 0.40\pm0.10$.}
    \label{f3}
\end{figure*}

\begin{align}  
    H(\Phi)=&\sum_{j=1}^{L}[-t(e^{-g}e^{-i \frac{\Phi}{L}j}\hat{b}^\dagger_{j+1}\hat{b}_j+e^{g}e^{i \frac{\Phi}{L}j}\hat{b}^\dagger_j \hat{b}_{j+1})\\ 
    &+U\hat{n}_{j}\hat{n}_{j+1}+\Delta_j \hat{n}_j].\nonumber
\end{align}
Based on this transformation, the winding number is defined, denoted as 
\begin{align}
    \omega_{\Phi} = \int _ { 0 } ^ { 2 \pi } \frac { d \Phi } { 2 \pi i } \partial _ { \Phi }  \{ H ( \Phi ) - E _ { B } \}
    \label{eq8}
\end{align}
Note that $E_B$ refers to a specified reference point rather than the ground state energy. Departing from the traditional bulk-edge correspondence found in Hermitian systems, $\omega_{\Phi}$ is used to compute the count of complex eigenenergy trajectory encirclements around the reference point $E_B$ as the phase $\Phi$ transitions from $0$ to $2\pi$. Here, ${\rm det}H(\Phi)/|{\rm det}H(0)|$ is employed to characterize the winding number around the reference point $E_B$ \cite{PhysRevB.82.073105}. As a result, the winding number $\omega_{\Phi}$ does not directly pertain to topological edge states but can be harnessed to unveil topological phase transitions.

Next, we examine spectral RC transition. We have plotted the eigenvalues at size $L=12$ under $\gamma=0.2$ and $\gamma=4.0$, as shown in FIG. \ref{f3}(a) and FIG. \ref{f3}(b) respectively. Since the TRS, all eigenvalues with imaginary parts are symmetrically distributed along the real axis. It can be seen that when $\gamma=0.2$, there are more eigenvalues with nonzero imaginary parts. However, when in a deeper MBL region, i.e., $\gamma=4.0$, almost all eigenvalues fall on the real axis. 

\begin{figure}[htbp]
    \centering
    \includegraphics[width=1\textwidth]{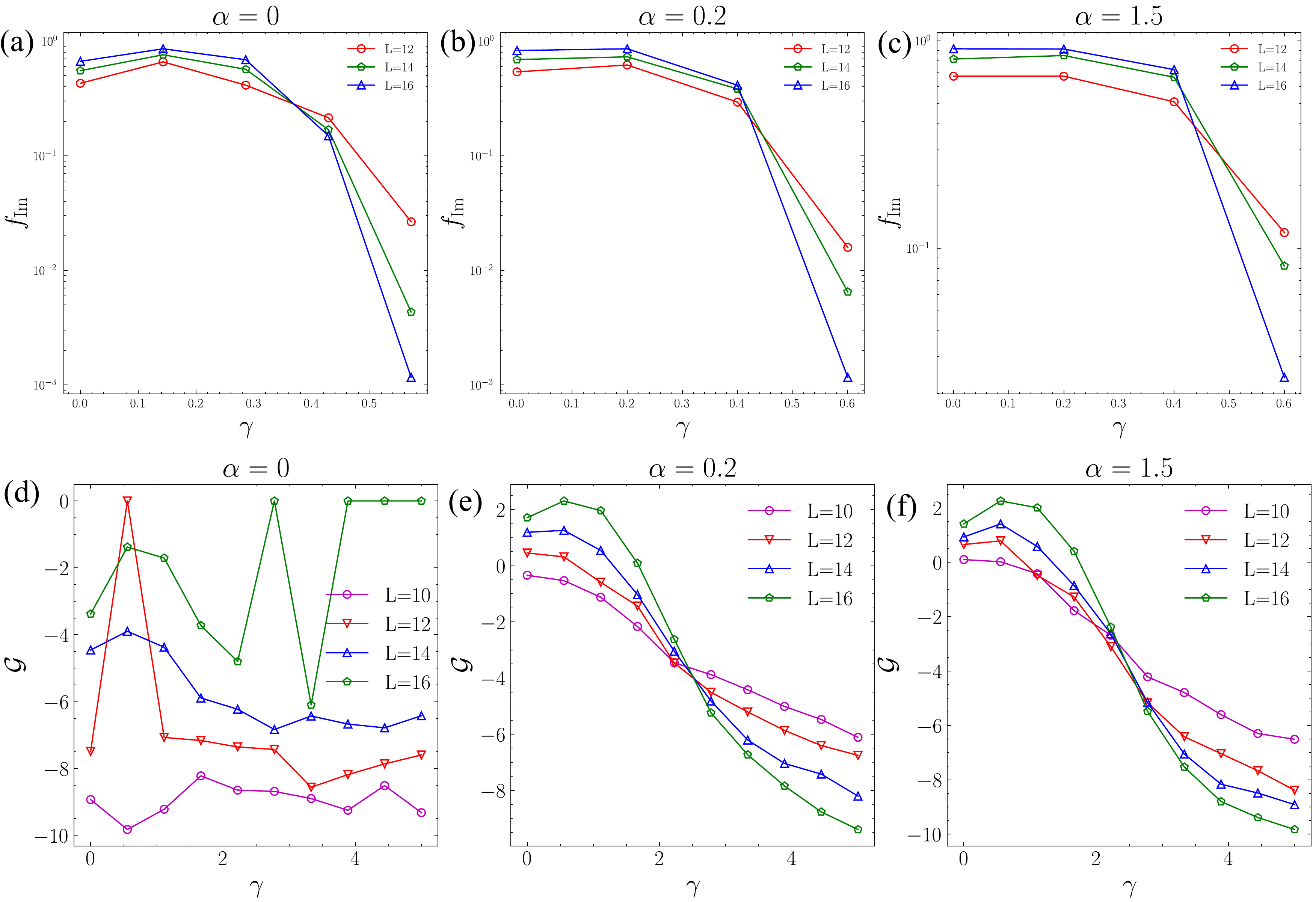}
    \caption{(Color online) Effects of tail curvature $\alpha$ on NH-SMBL system's phase transitions. When $\alpha=0$, as depicted in (a), the system exhibits a spectral RC transition. While the MBL transition remains undetectable due to unbroken translation symmetry (d). When $\alpha$ increases to $0.2$, as shown in (b), the spectral RC transition persists and the MBL transition begins to appear as translation symmetry is weakly broken (e). Further increasing $\alpha$ to $1.5$, as represented in (c), shows a minor shift in the critical points of both the spectral RC transition and the MBL transition (f). This underscores the role of weakly broken translational symmetry, induced by tail curvature, in achieving the robustness of SMBL.}
    \label{f6}
\end{figure}

We also identify the winding number of the energy spectrum and plotted the winding number ${\rm det}H(\Phi)/|{\rm det}H(0)|$ over the complex energy spectrum encircling the origin point $E_B=0$. The value of $\omega_{\Phi}$ in Eq.(\ref{eq8}) is derived from the number of times the loop encircles the origin. FIG. \ref{f3}(a) and (b) illustrates the loop of complex energy winding around the origin point $E_B$. In FIG. \ref{f3}(a), we depict the scenario when $\gamma=0.3$, while FIG. \ref{f3}(b) presents the case for $\gamma=0.45$. Evidently, the winding number around $E_B$ diminishes as $\gamma$ increases, ultimately reaching zero. Further, we have tallied the winding number for system sizes $L$ of 8, 10, 12, and 14, as demonstrated in FIG. \ref{f3}(d). It reveals a phase transition from a nonzero to zero winding number as the Stark gradient potential escalates. The critical point progressively stabilizes at $\gamma^{T}_c\approx 0.40\pm0.10$ with increasing system size, eventually reaching a plateau at $L=14$. The critical point for topological phase transition in interacting NH-TRS system under Stark potential is still considerably detached from that of the MBL phase transition, possibly linked to the complex topological characteristics inherent in non-Hermitian interacting systems. But the critical point of spectral RC transition is closely approached.

\section{Tail curvature of Stark potential}
\label{sec:s3}
The most important function of the tail curvature term in the Stark potential is to weakly break the translational invariance in order to achieve a degeneracy removal effect. We have computed the variation of eigenvalues nonzero ratio $f_{\rm Im}$ and stability of eigenstate $\mathcal{G}$ with respect to the system size $L$ when the tail curvature parameter $\alpha=0, 0.2, 1.5$ cases. In the case of a pure linear Stark potential $\alpha=0$, each eigenstate is localized around a site with an inverse localization length that remains energy-independent, even deep into the localized phase. This results in a spectrum forming an ordered ladder. Therefore, we can weakly break translation invariance by introducing a small tail curvature or adding small disorder to resolve degeneracy and achieve robust localization. In this paper, we chose to weakly break translational symmetry by adding a quadratic tail $\alpha$. The term $1/L^2$  ensures that the linear term plays the dominant role in the thermodynamic limit. As depicted in FIG.~\ref{f6}, for the linear case ($\gamma=0$), an RC transition is apparent (a), yet $\mathcal{G}$ does not exhibit any noticeable crossover behavior across different system sizes $L$ (b). In FIG.~\ref{f6}, under the condition of $\alpha=0.2$, the RC transition persists, while the system starts to exhibit an MBL transition as the translation symmetry is weakly broken [see FIG.~\ref{f6}(b) and (e)]. As $\alpha$ increases further to $1.5$, both the RC transition and the MBL transition see a minor shift in their respective critical points [see FIG.~\ref{f6}(c) and (f)]. Therefore, $\alpha=0.5$ is not a special selection for our system. Therefore, the primary role of the tail curvature is to weakly break translational symmetry, thereby achieving the robustness of SMBL by lifting the degeneracy.

\section{SPECTRAL RC TRANSITIONS AND MBL PHASE TRANSITIONS UNDER OBCs}
\label{sec:S2}
\begin{figure}[htbp]
    \centering
    \includegraphics[width=1\textwidth]{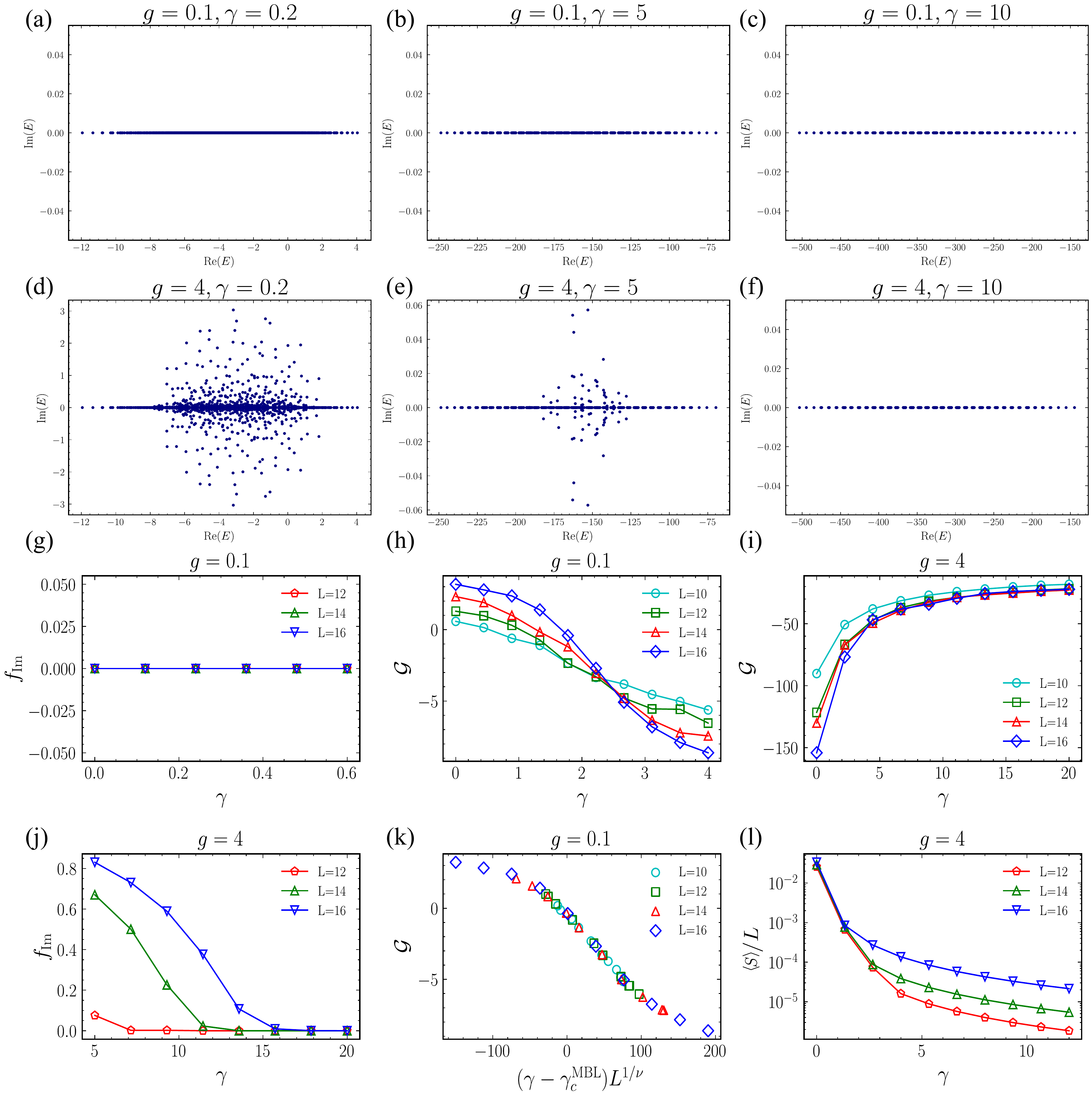}
    \caption{(Color online) Spectral RC transitions and MBL phase transitions under OBCs. (a)-(c) depict the distribution of eigenenergies under various conditions defined by parameters $g$ and $\gamma$. Subfigures (g) and (j) show $f_{\rm Im}$ for $g=0.1$ and $g=4$, respectively. Subfigures (h) and (k) present $\mathcal{G}$ at $g=0.1$, exhibiting a critical point and critical exponent estimated to be $\gamma^{\rm MBL}_{c}\approx 2.13\pm0.10$ and $\nu\approx 0.66\pm0.14$, respectively. Finally, subfigures (i) and (l) display $\mathcal{G}$ and $\langle S \rangle/L$ respectively at $g=4$. All the aforementioned situations occur under the condition of $U=1$.}
    \label{s5}
\end{figure}

In this section, we consider Eq. [\ref{eq1}], with the boundary condition changed to OBC. The Hamiltonian is as follows:

\begin{align}
    \hat{H}= \sum _{j=1}^{L-1}\left[ -t(e^{-g }\hat{b}^{\dagger}_{j+1}\hat{b}_{j}+e^{g}\hat{b}^{\dagger}_{j}\hat{b}_{j+1})+U \hat{n}_{j}\hat{n}_{j+1}\right] + \sum_{j=1}^{L} \Delta _{j}\hat{n}_{j} .
\end{align}
The meanings of the parameters symbolized in this equation remain unchanged. In previous reports concerning disordered, non-Hermitian systems, it has been observed that spectral RC transitions occur simultaneously in both PBCs and OBCs \cite{PhysRevB.106.064208}. This leads us to question whether the speactral RC transitions and MBL phase transitions in non-Hermitian SMBL systems with OBCs could also demonstrate such robustness. Here, we have conducted numerical calculations under OBCs, revealing that in the context of OBCs, spectral RC transitions necessitate a higher non-reciprocal strength $g$ and Stark potential strength $\gamma$ to persist. As depicted in FIG.~\ref{s5}(a)-(c), under the condition of $g=0.1$, complex energy levels do not emerge for $\gamma$ ranging from 0 to 10. However, when $g=4$, spectral RC transitions are discernible, as shown in FIG.~\ref{s5}(d)-(f), FIG.~\ref{s5}(g) and FIG.~\ref{s5}(j). In the case of the MBL phase transition at $g=0.1$, both the critical point and the critical exponent remain virtually unchanged, as demonstrated in FIG.~\ref{s5}(h) and FIG.~\ref{s5}(k). When the parameter is adjusted to $g=4$, the MBL phase transition disappears, as can be seen in FIG.~\ref{s5}(i) and FIG.~\ref{s5}(l).

\end{document}